\documentclass[aps,twocolumn,prl,showpacs,amsmath,amssymb,10pt,showkeys]{revtex4-2}

\usepackage{graphicx}
\usepackage{scrextend}
\usepackage{manfnt,pifont}
\usepackage{hyperref}
\usepackage{url}
\usepackage{lipsum}
\usepackage[font=small]{subcaption}
\usepackage[font=small]{caption}
\usepackage{comment}

\usepackage[normalem]{ulem}

\usepackage{url}
\usepackage[outline]{contour}
\usepackage{textgreek}

\usepackage{tikz}
\usetikzlibrary{shapes}

\usepackage{CJKutf8}
\usepackage{enumitem}
\setlist[itemize]{leftmargin=*}

\newcommand\wh\widehat

\makeatletter
\DeclareFontFamily{OMX}{MnSymbolE}{}
\DeclareSymbolFont{MnLargeSymbols}{OMX}{MnSymbolE}{m}{n}
\SetSymbolFont{MnLargeSymbols}{bold}{OMX}{MnSymbolE}{b}{n}
\DeclareFontShape{OMX}{MnSymbolE}{m}{n}{
    <-6>  MnSymbolE5
   <6-7>  MnSymbolE6
   <7-8>  MnSymbolE7
   <8-9>  MnSymbolE8
   <9-10> MnSymbolE9
  <10-12> MnSymbolE10
  <12->   MnSymbolE12
}{}
\DeclareFontShape{OMX}{MnSymbolE}{b}{n}{
    <-6>  MnSymbolE-Bold5
   <6-7>  MnSymbolE-Bold6
   <7-8>  MnSymbolE-Bold7
   <8-9>  MnSymbolE-Bold8
   <9-10> MnSymbolE-Bold9
  <10-12> MnSymbolE-Bold10
  <12->   MnSymbolE-Bold12
}{}

\let\llangle\@undefined
\let\rrangle\@undefined
\DeclareMathDelimiter{\llangle}{\mathopen}%
                     {MnLargeSymbols}{'164}{MnLargeSymbols}{'164}
\DeclareMathDelimiter{\rrangle}{\mathclose}%
                     {MnLargeSymbols}{'171}{MnLargeSymbols}{'171}
\makeatother

\usepackage{eso-pic}

\begin{document}
\begin{CJK*}{UTF8}{gbsn}

\title{The Roaming Bethe Roots: An Effective Bethe Ansatz Beyond Integrability}

\author{Wenlong Zhao ({\CJKfamily{gbsn}赵文龙})$^{a}$}
\author{Yunfeng Jiang ({\CJKfamily{gbsn}江云峰})$^{b,e}$}\altaffiliation{Corresponding authors}
\author{Rui-Dong Zhu ({\CJKfamily{gbsn}朱睿东})$^{a,c,d}$}\altaffiliation{Corresponding authors}
\affiliation{$^a$School of Physical Science and Technology \& Institute for Quantum Science, Soochow University, Suzhou 215006, China}
\affiliation{$^b$School of Physics \& Shing-Tung Yau Center, Southeast University, Nanjing 211189, P. R. China}
\affiliation{$^c$Suzhou Municipal Key Laboratory of Quantum-Intelligence Fusion Computing, Suzhou 215000, China}
\affiliation{$^d$Jiangsu Key Laboratory of Frontier Material Physics and Devices,\\ Soochow University, Suzhou 215006, China}
\affiliation{$^{e}$Peng Huanwu Center for Fundamental Theory, Hefei, Anhui 230026, China}
\keywords{Bethe ansatz; Integrability breaking; Spin chain}

\begin{abstract}
We propose an effective Bethe ansatz (EBA) for solving quantum many-body systems near an integrable point. Our approach retains the functional form of the Bethe wave function while renormalizing the Bethe roots to account for integrability-breaking interactions. These effective roots are determined by minimizing physically motivated cost functions. The resulting off-shell Bethe states serve as approximate eigenstates of the non-integrable models. We assess the quality of the approximation using various physical observables, including the energy eigenvalue, state fidelity, and bipartite entanglement entropy. Our tests show that for models with weak integrability-breaking, the effective Bethe ansatz provides a high-quality approximation to the exact eigenstates over a wide range of deformation parameters. In contrast, for models with strong integrability-breaking interactions, the efficacy of the effective Bethe ansatz degrades relatively quickly as the deformation parameter increases. The efficacy of the method thus offers a useful probe for characterizing the strength of integrability breaking. Within its regime of accuracy, it also provides a new representation of the eigenstates of nearly integrable models, enabling one to exploit the algebraic structure inherited from integrability.

\end{abstract}

\maketitle

\noindent{\bf Introduction.} Integrable models are special many-body systems that are exactly solvable. Their solvability originates from an extensive set of conserved quantities, which impose strong constraints on the dynamics and lead to many unique physical properties. A fundamental question in integrable models is what happens when integrability is broken. In classical systems, the Kolmogorov–Arnold–Moser theorem \cite{Kolmogorov1954OnCO,Arn63,Mos62} shows that weak integrability-breaking perturbations preserve most phase space tori; as the perturbation increases, the tori gradually break down, leading to chaos. In the quantum regime, systems near integrable points also exhibit properties reminiscent of integrability, such as approximate conservation laws \cite{Bargheer:2008jt,Bargheer:2009xy,2015PhRvB..92s5121M,2020arXiv200701715M,2024PhRvB.110n4309V,PhysRevX.5.041043}, prethermalization \cite{Berges:2004ce,2011PhRvB..84e4304K,Langen:2016vdb,Mallayya:2019uqw,Moeckel:2008glt} and anomalous transport \cite{Znidaric:2011qvy,2022ScPP1313M,2007PhRvB76x5108J,2013PhRvB..88k5126H,2015PhRvB..91k5130K,2016PhRvL.116a7202S,Durnin:2020kcg,Bastianello2021}. These observations may be interpreted as evidence that the influence of integrability or similar physical properties does not cease abruptly under deformation, but instead diminishes continuously with growing perturbation strength.\par

This picture implies that sufficiently close to an integrable point, the influence of integrability remains dominant. It provides useful intuition about the physical properties one can expect for models near integrability. At the same time, it offers hints for how to actually solve such models. Integrability is a powerful tool when applicable. Near an integrable point, one can pursue a hybrid method that exploits the integrability structure while incorporating integrability-breaking effects in a controlled manner. Indeed, one may treat the integrability-breaking part as a perturbation and apply standard perturbation theory \cite{Zamolodchikov:1989fp,Delfino:1996xp}, or adopt the Hamiltonian truncation approach \cite{YUROV-ZAMOLODCHIKOV,Yurov:1989yu,Fonseca:2001dc}. 

In this work, we pursue a different route to studying models near integrability, based on the scattering picture. A key idea behind integrable models is factorized scattering, where the system can be viewed as a collection of stable particles interacting in a factorizable way. What happens to factorized scattering when integrability is broken? It is reasonable to assume that sufficiently close to the integrable point, the factorized scattering picture remains valid, but certain parameters must be renormalized. This is analogous to Fermi liquid theory, where interactions renormalize parameters such as the effective mass. When and how does the factorized scattering picture break down as integrability-breaking interactions are turned on? To investigate this question concretely, in this work, we focus on models solvable by the Bethe ansatz.

In the Bethe ansatz, the wave function is built upon the assumption of factorized scattering \cite{Bethe:1931hc,PhysRev.130.1605,Yang:1967bm,Baxter:1982zz}. In finite volume, the rapidities of pseudoparticles are quantized by the Bethe ansatz equations (BAE). When integrability is broken, the Bethe ansatz no longer yields exact solutions. However, we could assume that sufficiently close to an integrable point, the functional form of the Bethe wave function remains valid while the Bethe roots are deformed. To determine these deformed roots, we introduce a numerical optimization approach: we define suitable cost functions that depend on the deformed roots and employ modern optimization algorithms to find the minima. This approach, using the Bethe ansatz as a variational ansatz in non-integrable models, restores the original spirit of the ansatz as a trial wave function and shall be called an \emph{effective Bethe ansatz} (EBA). A similar idea has been explored in to solve the (integrable) XXZ \cite{Nepomechie:2021jak,Raveh:2024cfh} and Lieb-Liniger models \cite{PhysRevB.101.045102}, and \cite{Claeys:2017ipj} employed the Richardson-Gaudin state as a variational ansatz for approximating near-integrable models.

%Taking advantage of the exact solvability of integrable models, one can develop methods to tackle nearly integrable systems by constructing suitable perturbed versions of integrability-based approaches. The regime of validity of such methods naturally depends on how quickly integrability is lost, which in turn is governed by the nature of the perturbation. Can integrability be leveraged to solve non-integrable models? How can one characterize the rate at which integrability is lost? Are there specific values of the perturbation parameter at which an abrupt transition occurs? Motivated by these questions, in this paper we investigate a family of quantum integrable systems solvable by the Bethe ansatz and study the effects of integrability breaking. (\textcolor{red}{YF: Maybe comment on GHD for nearly integrable systems ?})\par

The resulting off-shell Bethe states with the deformed Bethe roots serve as an approximation of the exact eigenstates. To assess the validity of the effective Bethe ansatz, we compare energies, fidelities, and bipartite entanglement entropies of the effective Bethe states with exact eigenstates. A clear distinction emerges between two types of model: under the deformation of \eqref{H-weak}, the approximation degrades much more slowly than under \eqref{H-strong}, remaining accurate over a broad parameter range. The rate of breakdown thus serves as a diagnostic for the strength of integrability breaking. Moreover, we found that near specific points of the parameter space, such as a level crossing point or a phase transition point, the approximation undergoes a sharp decline, suggesting that the effective Bethe ansatz can also be a useful probe for quantum criticality.

\vspace{0.5cm}

\noindent{\bf Bethe ansatz as an ansatz.} We consider a Hamiltonian of the form
\begin{align}
H(\lambda)=H_0+\lambda H_1,
\end{align}
where $H_0$ is an integrable spin chain solvable by the Bethe ansatz. We denote the eigenstates by $|\{u_j\}\rangle$, which depends on a set of rapidities $\{u_j\}$ that satisfy the BAE. The Hamiltonian is deformed by $\lambda H_1$, where $\lambda$ controls the strength of the deformation. The deformation may preserve integrability or, more generally, break it. Our central assumption is that, for sufficiently small $\lambda$, the Bethe state remains a good approximate eigenstate of $H(\lambda)$, as long as $\{u_j\}$ are chosen properly. The range of validity for $\lambda$ depends on the nature of the deformation operator $H_1$.

We distinguish two classes of deformations \footnote{So far, a rigorous definition of weak and strong integrability-breaking perturbations is still lacking. In fact, for a sufficiently large deformation parameter, all perturbations lead to strong integrability breaking. In this context, the distinction between weak and strong should be understood qualitatively. For the same deformation parameter, some perturbations break integrability more rapidly than others.}: \emph{weak} and \emph{strong} integrability breaking. The two types exhibit different physical properties. Weakly integrability-breaking models exhibit approximate conservation laws \cite{Bargheer:2008jt,Bargheer:2009xy,2020arXiv200701715M,Surace:2023wqq,Pozsgay:2019ekd,Marchetto:2019yyt,Doyon:2021tzy} and display behavior close to that of integrable systems, such as anomalously large heat conductivity \cite{2006PhRvL..96f7202J} and prethermalization \cite{Mallayya:2019uqw,Mallayya:2021fdg,Mori:2017qhg}. In contrast, strongly integrability-breaking perturbations tend to produce generic non-integrable behavior, including chaotic level statistics \cite{PhysRevB.69.054403,Santos:2010iji,Brandino:2010sv}, particle decay and production \cite{2014PhRvB..90q4406R,Delfino:1996xp}, soliton confinement \cite{McCoy:1978ta,Fonseca:2006au,2017NatPh..13..246K,2024PhRvB.109a4411R,Lagnese:2021hjt,Gao:2025mcg,Litvinov:2025geb} and false vacuum decay \cite{Coleman:1977py,Callan:1977pt,Lagnese:2021grb,Szasz-Schagrin:2022wkk}. It is therefore natural to expect that the effective Bethe ansatz will remain accurate for weakly integrability-broken models over a relatively wide parameter window, whereas for strongly-broken integrability the approximation will deteriorate rapidly, remaining useful only for very small $\lambda$. 

Assuming the Bethe state provides a good approximation in the appropriate regime, the deformed Bethe roots can be obtained by treating the off-shell Bethe state as a variational ansatz with parameters $\{u_j\}$. We optimize these parameters by minimizing a suitable functional. Specifically, for the ground state, we minimize
\begin{align}
\mathcal{H}_0(\lambda;\{u_j\})\equiv\frac{\langle\{u_j\}|H(\lambda)|\{u_j\}\rangle}{\langle\{u_j\}|\{u_j\}\rangle}.\label{grd-cost}
\end{align}
The resulting rapidities are denoted by $\{\bar{u}_j\}$. A convenient starting point for the optimization is provided by the original Bethe roots of the undeformed integrable spin chain. The resulting normalized ground state is then $|\psi_0\rangle = |\{\bar{u}_j\}\rangle/\sqrt{\langle\{\bar{u}_j\}|\{\bar{u}_j\}\rangle}$.

For excited states, we minimize a modified functional,
\begin{align}
\mathcal{H}_1(\lambda;\{u_j\}) = \mathcal{H}_0(\lambda;\{u_j\}) + K_1\frac{|\langle\{u_j\}|\psi_0\rangle|^2}{\langle\{u_j\}|\{u_j\}\rangle},
\end{align}
where $K_1 \gg 1$ is a large constant that effectively enforces orthogonality to $|\psi_0\rangle$. In general, this condition alone does not uniquely select the first excited state; the minimization may converge to other excited states. Lower-lying excitations can be singled out by initializing the search near the corresponding undeformed Bethe roots. The resulting rapidities are denoted by $\{\bar{u}_j^{(1)}\}$, and the normalized excited state is $|\psi_1\rangle = |\{\bar{u}_j^{(1)}\}\rangle/\sqrt{\langle\{\bar{u}_j^{(1)}\}|\{\bar{u}_j^{(1)}\}\rangle}$.

This procedure can be iterated by defining
\begin{align}
\mathcal{H}_n(\lambda;\{u_j\}) = \mathcal{H}_{n-1}(\lambda;\{u_j\}) + K_n\frac{|\langle\{u_j\}|\psi_{n-1}\rangle|^2}{{\langle\{u_j\}|\{u_j\}\rangle}},\label{ext-cost}
\end{align}
with $K_n\gg 1$. Minimizing $\mathcal{H}_n$ yields rapidities $\{\bar{u}_j^{(n)}\}$ for the $n$-th excited state, giving the approximate eigenstate $|\psi_n\rangle = |\{\bar{u}_j^{(n)}\}\rangle/\sqrt{\langle\{\bar{u}_j^{(n)}\}|\{\bar{u}_j^{(n)}\}\rangle}$.

We note that when the target Hamiltonian $H(\lambda)$ possesses degenerate energy levels, the cost function \eqref{grd-cost} or \eqref{ext-cost} will have multiple distinct optima. In such cases, the optimized Bethe roots are not unique: different initial guesses for the effective Bethe roots lead to different local minima, each corresponding to a particular linear combination within the degenerate subspace.

\vspace{0.5cm}

\noindent{\bf Models and Results.} We now test our proposal on concrete models. We take $H_0$ to be the periodic Heisenberg XXX spin chain,
\begin{align}
H_0=H_{\text{XXX}}=\sum_{n=1}^L\vec{\sigma}_n\cdot\vec{\sigma}_{n+1}
\end{align}
where $\vec{\sigma}_n=(\sigma_n^x,\sigma_n^y,\sigma_n^z)$ are the Pauli matrices. The model is solved exactly by the algebraic Bethe ansatz (ABA), which we briefly review in SM. Within the ABA framework, a Bethe state takes the form
\begin{align}
|\{u_j\}\rangle=\mathbf{B}(u_1)\ldots \mathbf{B}(u_M)|\Omega\rangle
\end{align}
where the operator $\mathbf{B}(u)$ is defined in Appendix and $|\Omega\rangle=|\uparrow^L\rangle$ denotes the ferromagnetic reference state. For the deformation term $H_1$, we consider two representative cases:
\begin{itemize}
%\item Integrable deformation
%\begin{align}
%H_1=\sum_{n=1}^L\sigma_n^z\sigma_{n+1}^z\,,
%\end{align}
%which yields the integrable XXZ spin chain.
\item A weak integrability-breaking deformation
\begin{align}
H_1=\sum_{n=1}^L\vec{\sigma}_n\cdot\vec{\sigma}_{n+2}, \label{H-weak}
\end{align}
corresponding to the $J_1$-$J_2$ model, and charges conserved up to ${\cal O}(\lambda^2)$ corrections were constructed in \cite{Surace:2023wqq}, demonstrating integrability at ${\cal O}(\lambda)$ (see also \cite{Serban:2013,Orlov:2023}). 
\item A strong integrability-breaking deformation
\begin{align}
H_1=\sum_{n=1}^L(-1)^n\sigma_n^z\label{H-strong}
\end{align}
which introduces a staggered magnetic field.
\end{itemize}

Notably, the weak integrability-breaking term \eqref{H-weak} preserves SU(2) symmetry and is an RG‑irrelevant perturbation \cite{Beria:2013hz} that leaves the spectrum gapless, whereas the strong breaking term \eqref{H-strong} breaks SU(2) and also translation symmetries, and it gives an RG‑relevant perturbation (positive RG eigenvalue) that opens a gap.

We analyze several quantities to assess the quality of the approximation. As a basic test, we first compare the energy expectation value $E^{\text{(EBA)}}=\langle\psi_{\text{EBA}}|H(\lambda)|\psi_{\text{EBA}}\rangle/\langle\psi_{\text{EBA}}|\psi_{\text{EBA}}\rangle$ obtained from the EBA with the exact energy from exact diagonalization (ED). For $L=4$, we use the EBA method to obtain the full spectrum of both models \footnote{If one uses random initial values for the optimization, it will be very hard to find the state that corresponds to the singular physical Bethe state in the XXX model (in our convention, $u_1=-i,u_2=0$). If one starts from the known solution to the BAE, then it will not be hard to identify it. }, and the largest relative error occurs for the antiferromagnetic ground state. To evaluate how close the EBA state is to the exact eigenstate, we compute the fidelity $F=|\langle\psi_{\text{ex}}|\psi_{\text{EBA}}\rangle|^2/(\langle\psi_{\text{EBA}}|\psi_{\text{EBA}}\rangle\langle\psi_{\text{ex}}|\psi_{\text{ex}}\rangle)$, where $|\psi_{\text{EBA}}\rangle$ is the EBA state and $|\psi_{\text{ex}}\rangle$ is the exact eigenstate. A fidelity close to 1 indicates that the state serves as a good approximation to the exact state. Finally, since entanglement structure plays an important role in characterizing many-body wave functions, we also compute the bipartite entanglement entropy for different subsystem sizes and compare the results for the EBA state with those for the exact states. All these quantities evolve in a synchronized manner. In addition to physical quantities, we examine how the effective Bethe roots are modified under continuous deformation. The results for the two types of deformations are presented below.

%bipartite entanglement entropy, and state fidelity $|\langle\psi_{\text{ex}}|\psi_{\text{EBA}}\rangle|^2/(\langle\psi_{\text{EBA}}|\psi_{\text{EBA}}\rangle\langle\psi_{\text{ex}}|\psi_{\text{ex}}\rangle)$ where $|\psi_{\text{EBA}}\rangle$ is the effective Bethe state and $|\psi_{\text{ex}}\rangle$ is the exact eigenstate obtained from exact diagonalization (ED).

\vspace{0.3cm}
\noindent{\it -Weak integrability-breaking.}
The energy for the ground state and the first excited state of different lengths are given in Table~\ref{tab:weak-energy}. We compared the result for $\lambda=0.1$ with lengths from $L=4$ to $L=10$, the error of the energy is within $0.1\%$ for the ground state and $0.3\%$ for the 1st excited state. We have also computed at other deformation parameters ranging from $0.1$ to $0.9$ and higher lengths (for ground state at $\lambda=0.1,\ 0.3$ up to $L=20$, with errors bounded by $0.11\%$ and $1.2\%$, respectively.).

\begin{figure*}[htbp]
  \centering
  \begin{subfigure}[b]{0.45\textwidth}
    \includegraphics[width=\textwidth]{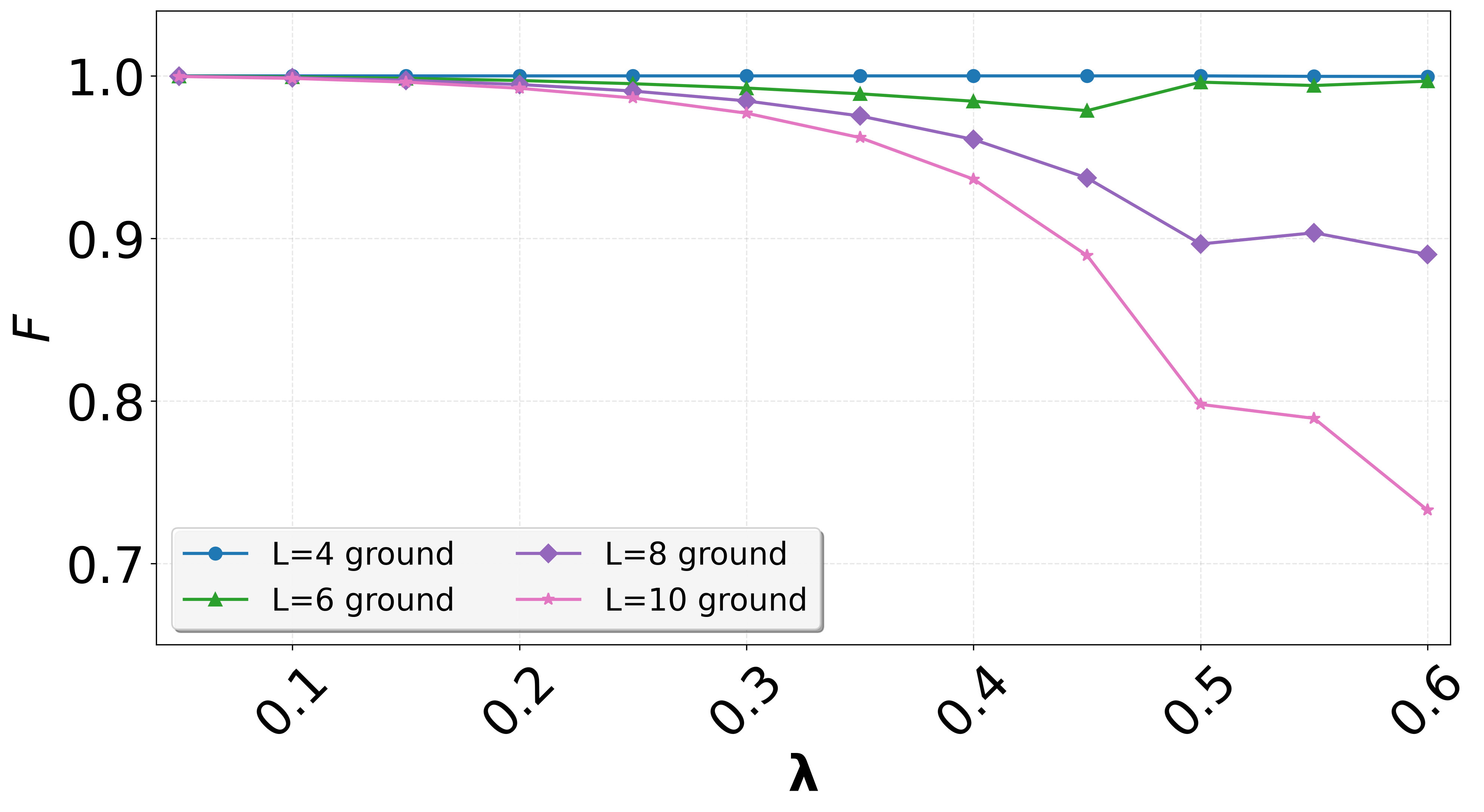}
    \caption{Ground-state fidelity in the model with weak integrability-breaking perturbation.}
    \label{fig:1a}
  \end{subfigure}
  \hfill
    \begin{subfigure}[b]{0.45\textwidth}
    \includegraphics[width=\textwidth]{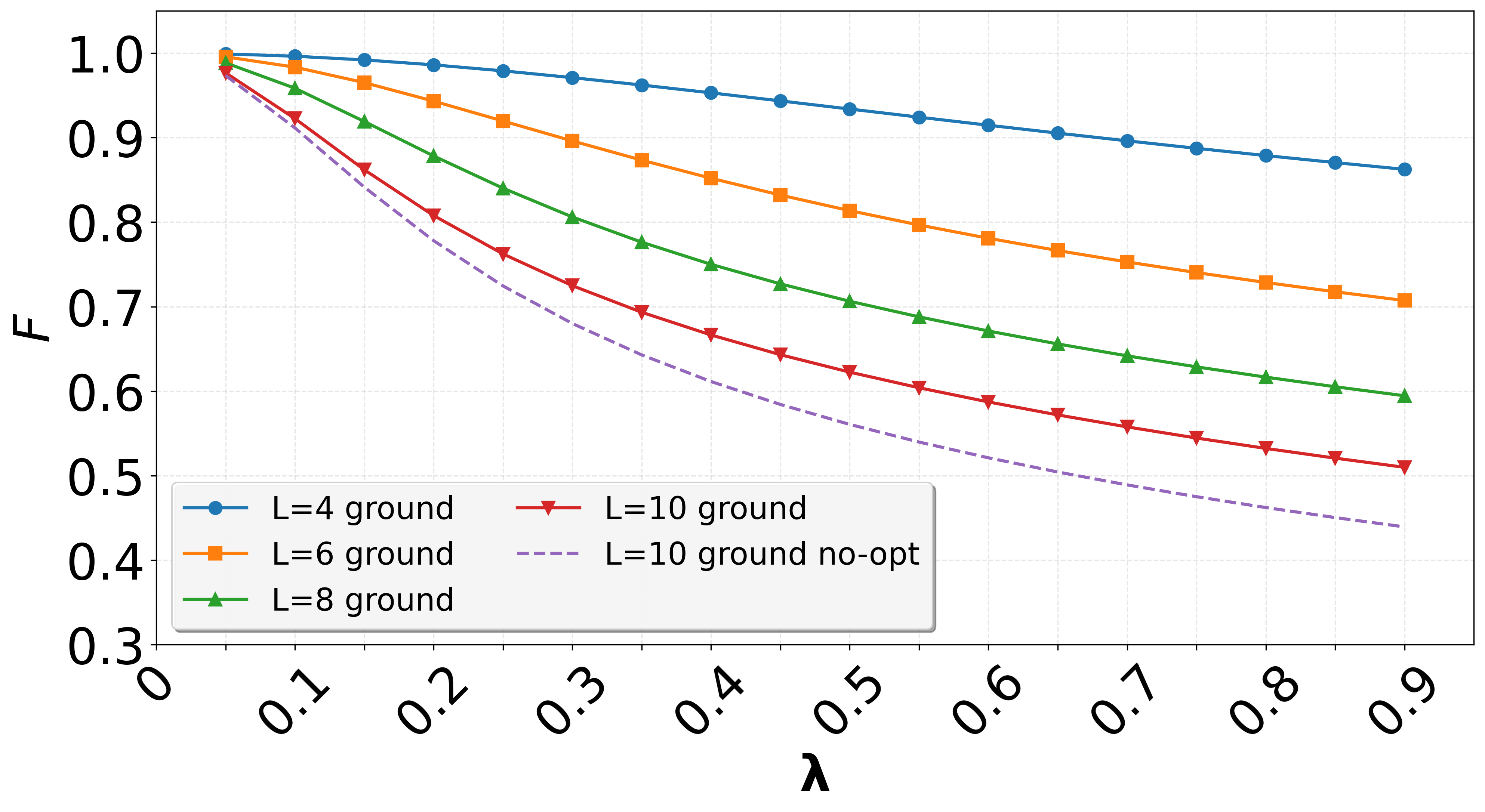}
    \caption{Ground-state fidelity in the model with strong integrability-breaking perturbation.}
    \label{fig:1b}
  \end{subfigure}
  \vspace{0.5cm}
    \begin{subfigure}[b]{0.45\textwidth}
    \includegraphics[width=\textwidth]{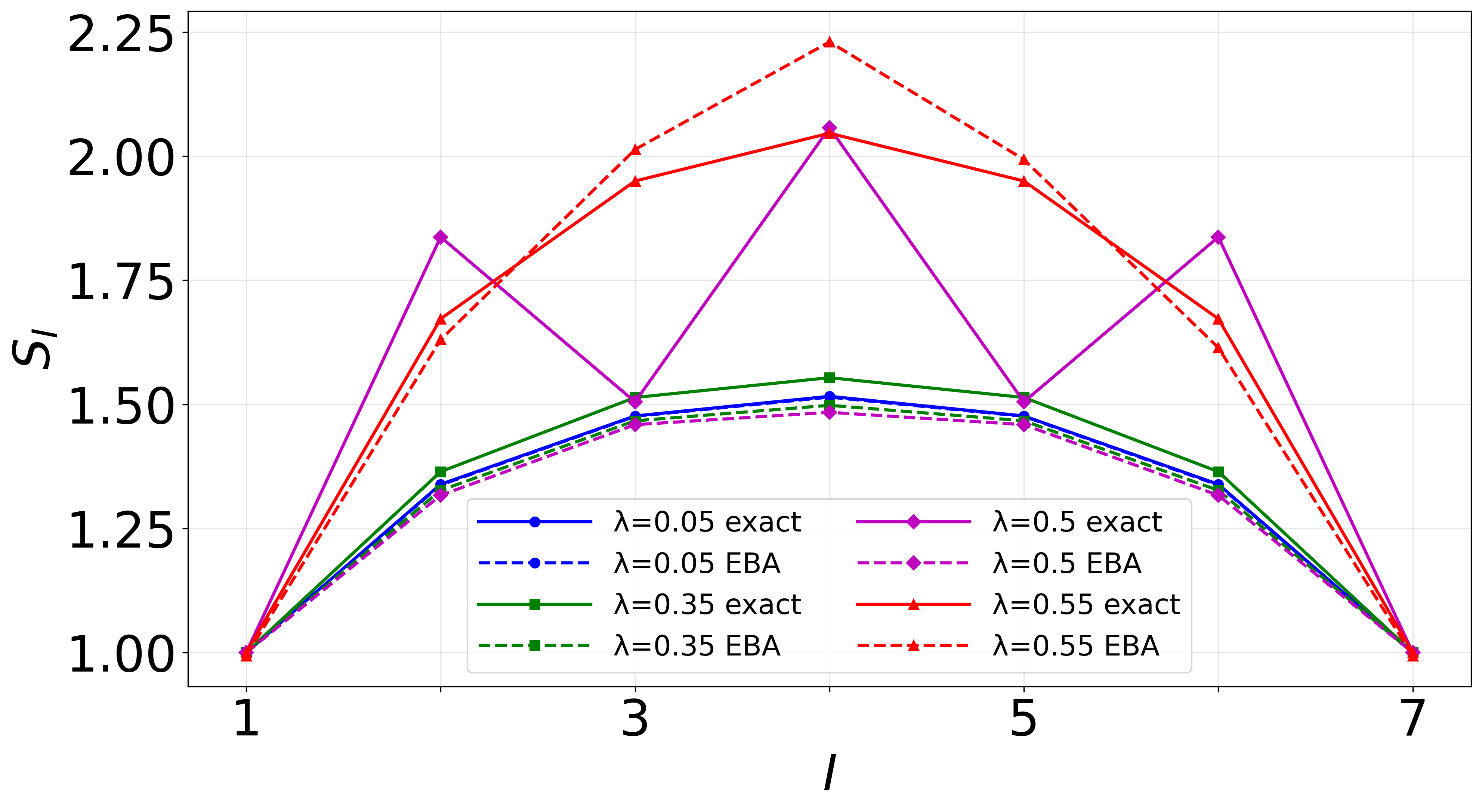}
    \caption{Bipartite entanglement entropy of $L=8$ ground state of weak integrability-breaking model. }
    \label{fig:1c}
  \end{subfigure}
  \hfill
  \begin{subfigure}[b]{0.45\textwidth}
    \includegraphics[width=\textwidth]{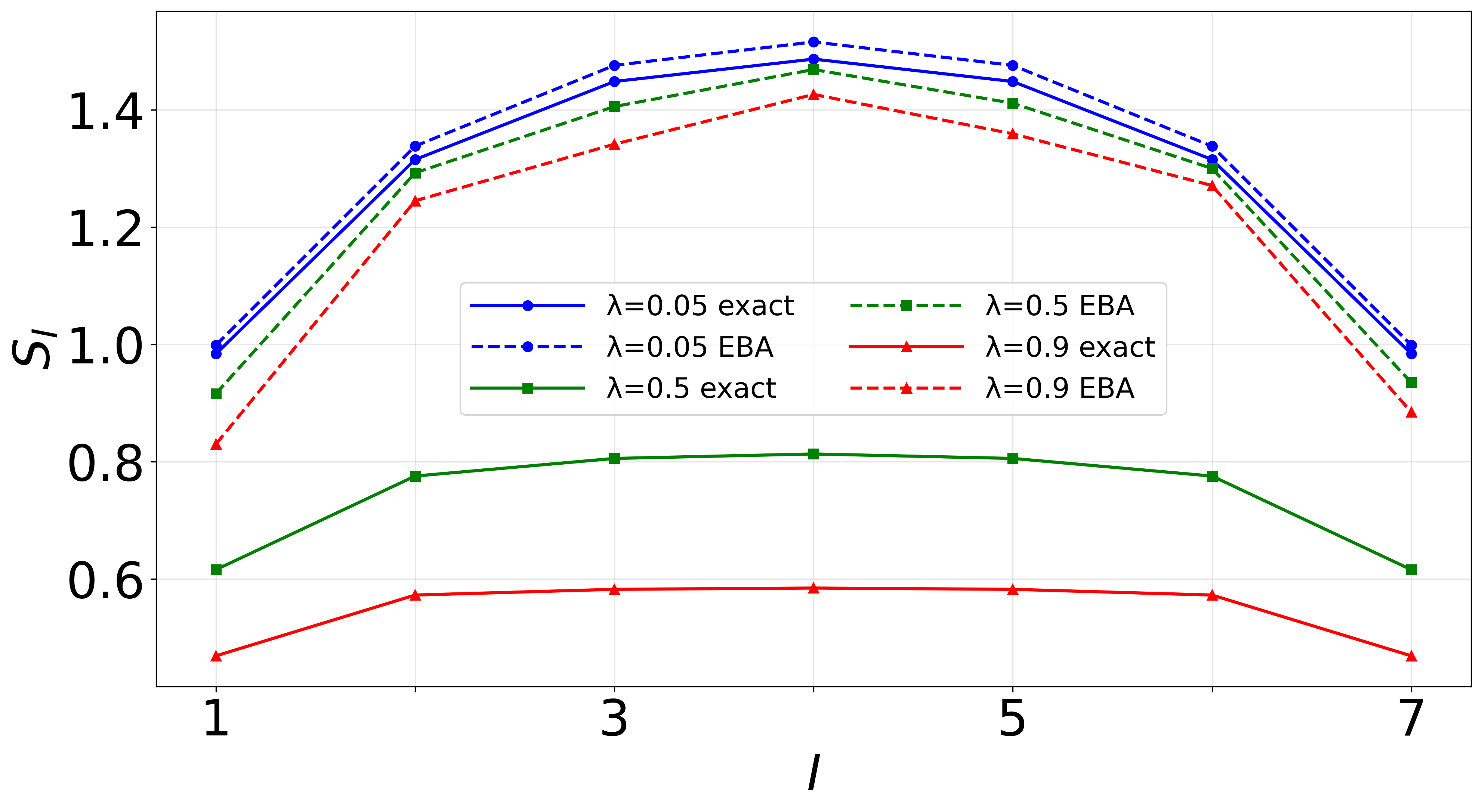}
    \caption{Bipartite entanglement entropy of $L=8$ ground state of strong integrability-breaking model.}
    \label{fig:1d}
    \end{subfigure}
  \caption{Fidelity and Entanglement Entropy computed in two models.}
  \label{fig:direct}
\end{figure*}

\begin{figure*}[htbp]
  \centering
  \begin{subfigure}[b]{0.45\textwidth}
    \includegraphics[scale=0.08]{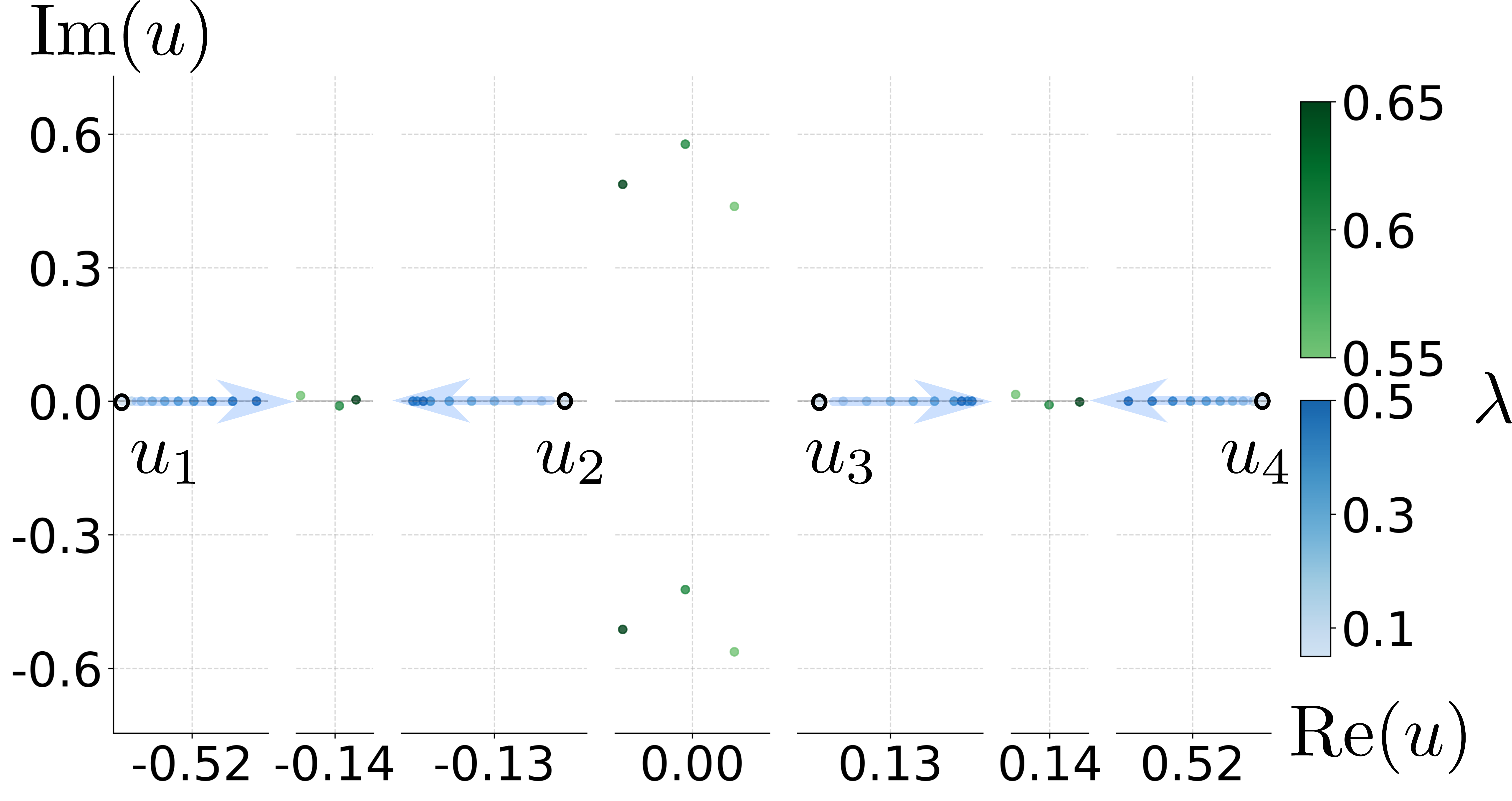}
    \caption{In weak integrability-breaking model}
    \label{fig:2a}
  \end{subfigure}
  \hfill
  \begin{subfigure}[b]{0.45\textwidth}
    \includegraphics[scale=0.078]{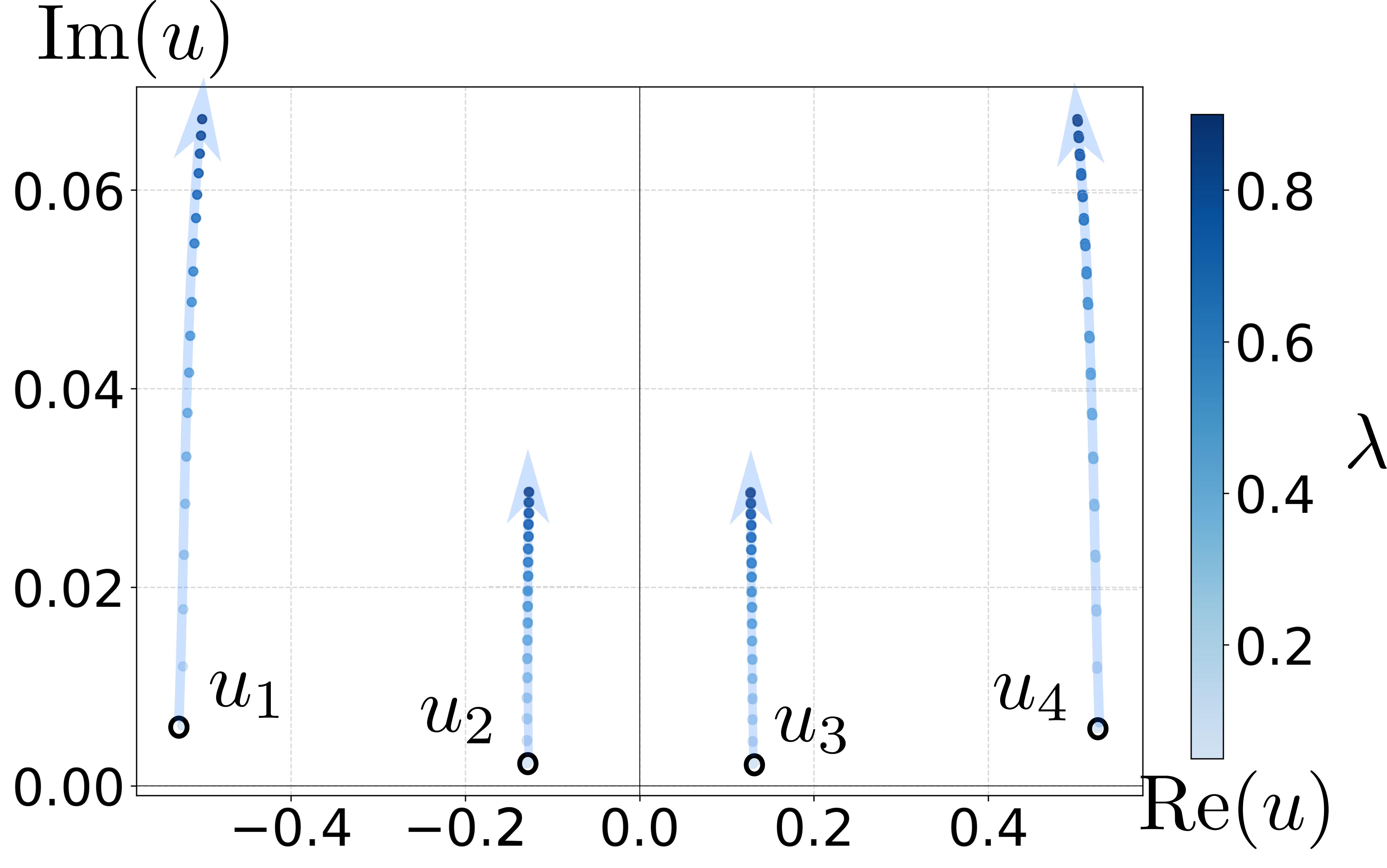}
    \caption{In strong integrability-breaking model}
    \label{fig:2b}
  \end{subfigure}
  \caption{Evolution of the effective Bethe roots of the ground state of two models with $L=8$. The black circles mark the positions of the Bethe roots corresponding to the undeformed ground state. The arrows indicate the directions that the corresponding Bethe roots move as the deformation parameter increases. For the weak integrability-breaking model, the effective Bethe roots changed abruptly at the MG point $\lambda=0.5$, indicated by the green dots in the figure. }
  \label{fig:u-pattern}
\end{figure*}

\begin{figure*}[htbp]
  \centering
  \begin{subfigure}[b]{0.45\textwidth}
    \includegraphics[width=\textwidth]{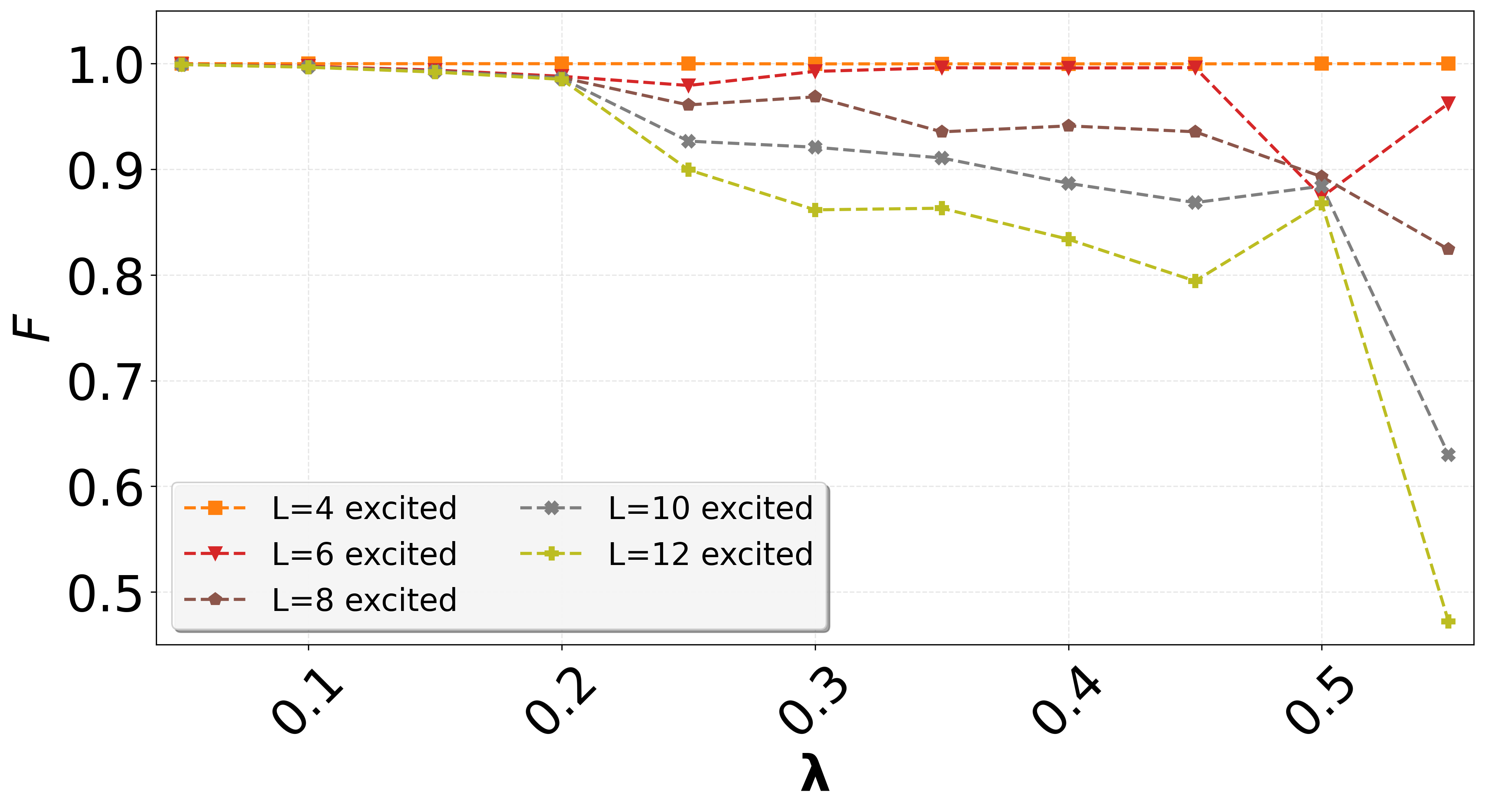}
    \caption{Fidelity of 1st excited state in weak integrability-breaking model computed from EBA.}
    \label{fig:3a}
  \end{subfigure}
  \hfill
  \begin{subfigure}[b]{0.45\textwidth}
    \includegraphics[width=\textwidth]{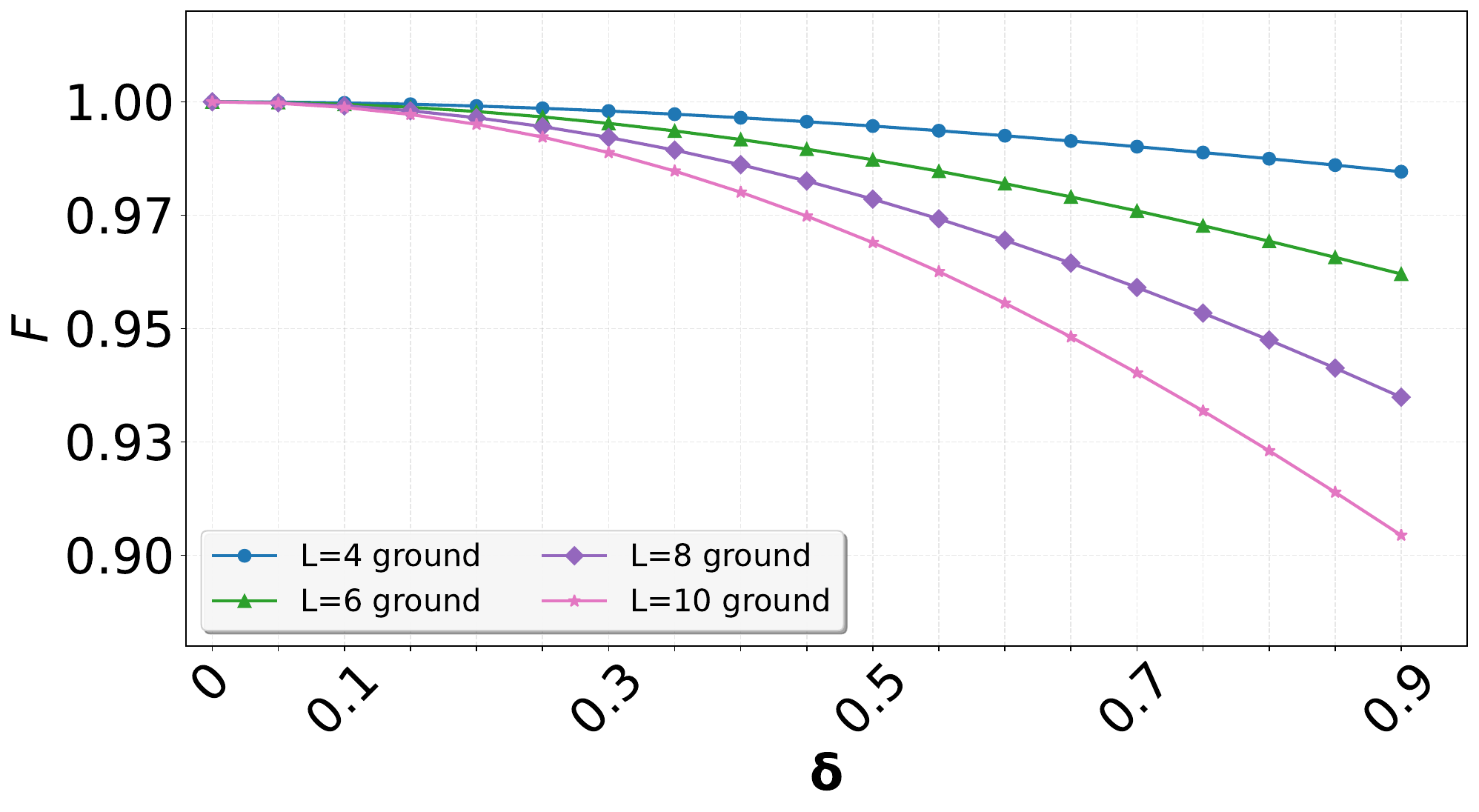}
    \caption{Ground-state fidelity in XXZ model computed from EBA, where $\delta=\Delta-1$}
    \label{fig:3b}
  \end{subfigure}
  \caption{Fidelity computed in the weak integrability-breaking model and the XXZ model.}
  \label{fig:xxz}
\end{figure*}

The fidelity of the ground state is shown in Fig. \ref{fig:1a}, which remains close to $1$ in the region $\lambda < 0.5$. At $\lambda=0.5$, the fidelity exhibits a sudden drop for all states. {The $\lambda = 0.5$ is known as the Majumdar-Ghosh (MG) point \cite{Majumdar:1969zmb}. In a finite chain, a level-crossing occurs at the MG point between the ground state and the 1st excited state.} 
%Below this point, the ground state lies in the $S^z_{\text{tot}} = 0$ sector, whereas above the transition it shifts to the $S^z_{\text{tot}} = 1$ sector. It is reflected in both the fidelity shown in Figure~\ref{fig:1a} and 
The entanglement entropy shown in Fig.~\ref{fig:1c}, where an abrupt change is also observed at $\lambda = 0.5$ \footnote{The ground states at the MG point $\lambda=0.5$ are twofold-degenerate and dimerized. The following states, 
\begin{equation*}
    |D_{A}\rangle=\otimes_{j=1}^{L/2} \lvert\mathrm{singlet}\rangle_{2j-1,2j},\quad |D_{B}\rangle=\otimes_{j=1}^{L/2} \lvert\mathrm{singlet}\rangle_{2j,2j+1}, 
\end{equation*}
are eigenstates at the MG point. A clear even-odd effect is evident in the bipartite entanglement entropy of these states. Taking one end of the region between the 1st and the $L$-th site,  the entanglement entropy alternates between $\log_22=1$ (when cutting through a singlet pair) and $0$ for $|D_A\rangle$, and between $\log_22=1$ and $2\log_22=2$ for $|D_B\rangle$. Figure \ref{fig:1c} gives the bipartite entanglement entropy of the state that has large overlap with the EBA state. It is a linear combination of $|D_A\rangle$ and $|D_B\rangle$ (which are not orthogonal to each other), and its entanglement entropy varies between $1$ and about $2$ with an even-odd effect. Unfortunately, the EBA does not capture such an effect at this special point}. The distribution of the effective Bethe roots also undergoes a sudden change in pattern when crossing this point, as illustrated in Fig.~\ref{fig:2a}. In the thermodynamic limit, the ground state and the 1st excited state in the finite chain merge into a twofold-degenerate ground state, and the system is known to lie in a dimerized phase \cite{Haldane1982}. The MG point is simply a special solvable point inside the dimerized phase, and the level-crossing observed does not lead to a quantum phase transition. There is indeed a phase transition point from a gapless spin-fluid phase to the gapped dimerized phase at $\lambda\approx 0.2411$ \cite{TI1987,1992PhLA..169..433O,LIU2011100}. It is indicated by a similar abrupt slope change in the fidelity curve and a similar sudden jump of effective Bethe roots of the 1st excited state around $\lambda\sim 0.25$ in our approach (see Figure \ref{fig:3a}; there is again a jump at the MG point). 
These results suggest that the effective Bethe ansatz may serve as a probe for potential quantum phase transitions, even in relatively short spin chains.

\begin{table}[h]
    \centering
    \begin{tabular}{l|c|c|c|c}
    \hline
    \hline
        $\quad L$ &  4 & 6 & 8 & 10\\
        \hline
        $E_0^{H_0}$  & $-8$ & $-11.2111$ & $-14.6044$ & $-18.0618$\\
        \hline
        $E_0^{\text{(ED)}}$  & $-7.6$ & $-10.7201$ & $-13.9896$ & $-17.3147$\\
        \hline
        $E_0^{\text{(EBA)}}$ & $-7.6$ & $-10.7119$ & $-13.9779$ & $-17.2996$\\
        \hline
        $\mathrm{Var}(H)/\Delta E^2$ & $0$ & $1.5\times10^{-2}$ & $3.1\times10^{-2}$ & $5.8\times10^{-2}$\\
        \hline
        $E_0^{\text{(no-opt)}}$ & $-7.6$ & $-10.7119$ & $-13.9779$ & $-17.2996$\\
        \hline
        $E_0^{{(\rm 1st-pt})}$  & $-7.6$ & $-10.7119$ & $-13.9779$ & $-17.2996$\\
        \hline
        $E_1^{H_0}$  & $-4$ & $-8.4721$ & $-12.5137$ & $-16.3688$\\
        \hline
        $E_1^{\text{(ED)}}$  & $-3.6$ & $-8.0448$ & $-11.9742$ & $-15.6963$\\
        \hline
        $E_1^{\text{(EBA)}}$ & $-3.6$ & $-8.0253$ & $-11.9531$ & $-15.6740$\\
        \hline
        $\mathrm{Var}(H)/\Delta E^2$ & $0$ & $2.1\times10^{-2}$ & $3.8\times10^{-2}$ & $6.4\times10^{-2}$\\
        \hline
        $E_1^{\text{(no-opt)}}$ & $-3.6$ & $-8.0249$ & $-11.9529$ & $-15.6739$\\
        \hline
        $E_1^{{(\rm 1st-pt})}$  & $-3.6$ & $-8.0249$ & $-11.9529$ & $-15.6739$\\
        \hline
    \end{tabular}
    \caption{The ground state energy ($M=L/2$) and the first-excited state energy ($M=L/2-1$) computed from ED, our effective Bethe ansatz (EBA) method, and the 1st-order perturbation theory for the weakly integrability-broken model at $\lambda=0.1$. For comparison, the exact energy levels of the antiferromagnetic ground state $E^{H_0}_0$ and the first excited state $E^{H_0}_1$ of the Heisenberg model $H_0$ are also listed. $E^{\text{(no-opt)}}$ stands for the energy expectation value of the XXX Bethe states (without optimization of Bethe roots). ${\rm Var}(H)/\Delta E^2$ measures eigenstate quality with $\Delta E$ denoting the energy difference between the lowest and next-to-lowest levels in the same sector as the target state.}
    \label{tab:weak-energy}
\end{table}

\noindent{\it -Strong integrability-breaking }
For comparison, in the model with a strong integrability-breaking deformation, the effectiveness of EBA is visibly reduced. The energy spectrum is given in Table~\ref{tab:strong-energy}, where we take the same value of deformation parameter $\lambda=0.1$. A mismatch already occurs for the first decimal digit for $L=10$ with $\sim 1\%$ error. The rate of breakdown of the EBA approximation thus serves as a diagnostic for the type of perturbations.

As shown in Figure~\ref{fig:1b}, the fidelity rapidly drops below $0.9$ once the perturbation is turned on to $\lambda \sim 0.1$, and the entanglement entropy similarly deviates from its exact value around the same point (see Figure~\ref{fig:1d}). This indicates that the integrability structure is more strongly disturbed under such a deformation. For $M=1$, one can easily understand it: the Bethe ansatz gives a plane wave, while the translational invariance is broken in this model.

We also note that the effect of optimization of Bethe roots is more evident in the strong integrability-breaking model (see Table \ref{tab:weak-energy}, \ref{tab:strong-energy}, $E^{({\rm EBA})}$ vs $E^{({\rm no-opt})}$). As expected, optimization becomes increasingly beneficial for larger deformation $\lambda$ and size $L$ (see Fig. \ref{fig:1b}, \ref{fig:eba_vs_L}, no-opt curves). In the weak integrability-breaking model, the optimization improves the fidelity of the ground state only for $\lambda\gtrsim0.4$, and becomes indispensable beyond the MG point $\lambda=0.5$, where the Bethe roots dive into the complex plane.

In perturbation theory, when the unperturbed spectrum contains degenerate levels, the correct approximation eigenstates of the perturbed system are generally given by linear combinations of the degenerate unperturbed states at the leading order. To handle such cases, the effective Bethe ansatz must be upgraded: instead of a single off-shell Bethe state $|\{u_j\}\rangle$, a superposition of multiple off-shell Bethe states $\sum_nc_n|\{u^{(n)}_j\}\rangle$ in general needs to be employed as the variational ansatz. Through the optimization process, the EBA algorithm then naturally constructs the correct linear combinations that lift the degeneracy.

For instance, in the simple case of $L = 4$ chain with a staggered magnetic field $\lambda = 0.1$,  our method identifies the perturbed eigenstates as a superposition of two degenerate states from the integrable XXX point ($|{{\bar{u}_1=\pm0.5}}\rangle$ with $E=0$), yielding 
\begin{equation}
\frac{|{{\bar{u}_1=0.5}}\rangle \pm \mathrm{i} |{{\bar{u}_1=-0.5}}\rangle}{\sqrt{2}},
\end{equation}
with energies $E = \mp 0.2$. This behavior is reminiscent of the Stark effect in conventional perturbation theory. The single Bethe states $|{{\bar{u}_1=\pm0.5}}\rangle$ are no longer eigenstates in the perturbed system. 

\begin{table}[h]
    \centering
    \begin{tabular}{l|c|c|c|c}
    \hline
    \hline
        $\quad L$ &  4 & 6 & 8 & 10\\
        \hline
        $E_0^{H_0}$  & $-8$ & $-11.2111$ & $-14.6044$ & $-18.0618$\\
        \hline
        $E_0^{\text{(ED)}}$  & $-8.0265$ & $-11.2753$ & $-14.7211$ & $-18.2433$\\
        \hline
        $E_0^{\text{(EBA)}}$ & $-8.0089$ & $-11.2226$ & $-14.6181$ & $-18.0774$\\
        \hline
        $\mathrm{Var}(H)/\Delta E^2$ & $5.7\times10^{-3}$ & $2.2\times10^{-2}$ & $5.3\times10^{-2}$ & $9.5\times10^{-2}$\\
        \hline
        $E_0^{\text{(no-opt)}}$ & $-8$ & $-11.2111$ & $-14.6044$ & $-18.0618$\\
        \hline
        $E_0^{{(\rm 1st-pt})}$  & $-8$ & $-11.2111$ & $-14.6044$ & $-18.0618$\\        
        \hline
        $E_1^{H_0}$  & $-4$ & $-8.4721$ & $-12.5137$ & $-16.3688$\\
        \hline
        $E_1^{\text{(ED)}}$  & $-4.0050$ & $-8.4851$ & $-12.5380$ & $-16.4080$\\
        \hline
        $E_1^{\text{(EBA)}}$ & $-4.0017$ & $-8.4747$ & $-12.5171$ & $-16.3729$\\
        \hline
        $\mathrm{Var}(H)/\Delta E^2$ & $1.3\times10^{-3}$ & $7.9\times10^{-3}$ & $2.1\times10^{-2}$ & $4.1\times10^{-2}$\\
        \hline
        $E_1^{\text{(no-opt)}}$ & $-4$ & $-8.4721$ & $-12.5137$ & $-16.3688$\\
        \hline
        $E_1^{({\rm 1st-pt})}$  & $-4$ & $-8.4721$ & $-12.5137$ & $-16.3688$\\
        \hline
    \end{tabular}
    \caption{The ground state energy ($M=L/2$) and the first-excited state energy ($M=L/2-1$) computed from ED, our effective Bethe ansatz (EBA) method, and the 1st-order perturbation theory for the strongly integrability-breaking model at $\lambda=0.1$. For comparison, the exact energy levels of the antiferromagnetic ground state $E^{H_0}_0$ and the first excited state $E^{H_0}_1$ of the Heisenberg model $H_0$ are also listed. The no-opt rows give the energies from the XXX Bethe states without optimization of Bethe roots.}
    \label{tab:strong-energy}
\end{table}

%\textcolor{red}{YF: Present more physical explanations for these two cases.}

%\vspace{0.5cm}

\noindent{\bf Improvements of the ansatz.} 
So far we have used the off-shell Bethe states of the XXX spin chain as an ansatz, where the tunable parameters are the effective Bethe roots. It is possible to improve the ansatz by introducing more tunable parameters while preserving integrability. In this section, we discuss a few natural generalizations of the ansatz.
\vspace{0.3cm}

\noindent{\it -More general $R$-matrix.} The off-shell Bethe ansatz of XXX model is constructed by using the $R$-matrix of the isotropic 6-vertex model. It is straightforward to generalize this construction by employing the anisotropic or $q$-deformed $R$-matrix of XXZ type. In this way, we can introduce $q$ as another tunable parameter.

Interestingly, one may use the XXX off-shell Bethe state to approximate an on-shell Bethe state of the XXZ spin chain. As the anisotropic parameter $\Delta=(q+q^{-1})/2$ deviates from 1, the fidelity decreases slowly and smoothly, as shown in Fig. \ref{fig:3b}. Naturally, if we include $q$ as an optimization parameter, we achieve a perfect match with the exact XXZ eigenstate.

\begin{figure*}[htbp]
  \centering
  \begin{subfigure}[b]{0.45\textwidth}
    \includegraphics[width=\textwidth]{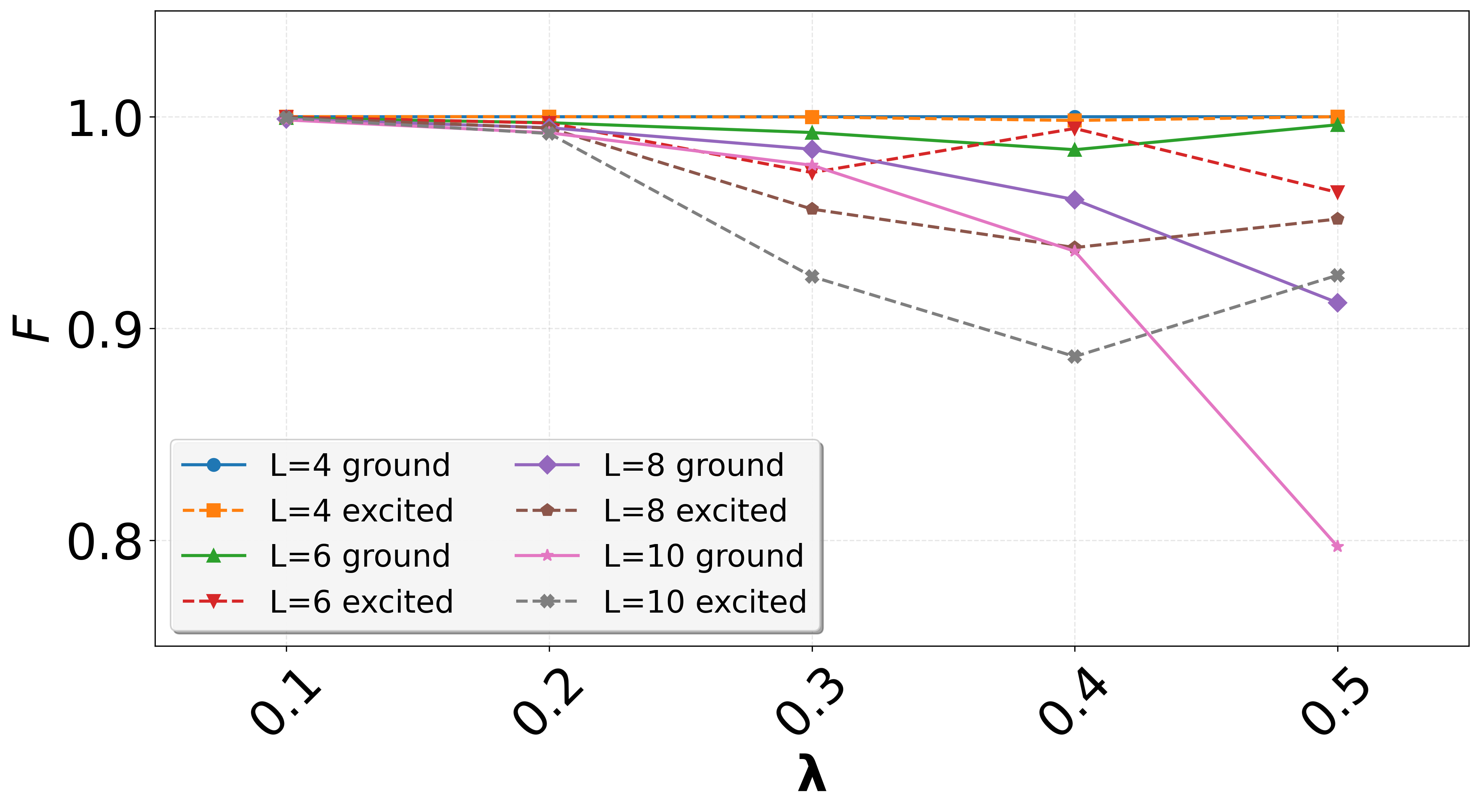}
    \caption{Fidelity in the model with weakly broken integrability}
    \label{fig:4a}
  \end{subfigure}
  \hfill
  \begin{subfigure}[b]{0.45\textwidth}
    \includegraphics[width=\textwidth]{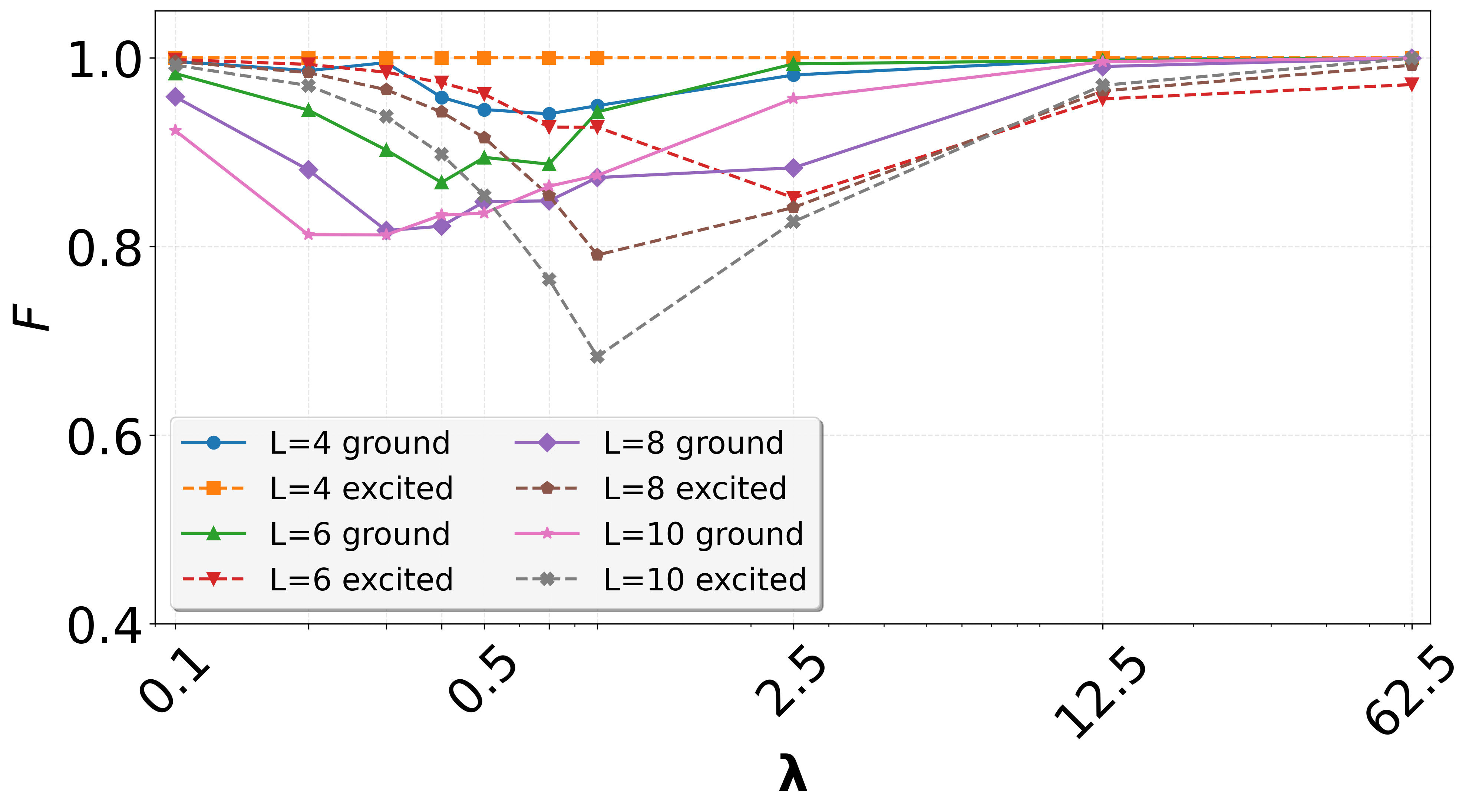}
    \caption{Fidelity in the model with strongly broken integrability}
    \label{fig:4b}
  \end{subfigure}
  \caption{Fidelity computed in two models based on an effective Bethe ansatz generated from the 8-vertex R-matrix.}
  \label{fig:8-vet}
\end{figure*}

In this example, the imperfection of the EBA originates from an inappropriate ansatz, instead of integrability breaking. This motivates us to adopt the following more general 8-vertex ansatz, which introduces two additional parameters:
\begin{equation*}
R(u;\beta,\gamma)=
\left(\begin{matrix}
u+\frac{\mathrm{i}(1+\beta)}{2} & & & \frac{\mathrm{i}(1-\gamma)}{2}\\
& u+\frac{\mathrm{i}(1-\beta)}{2} & \mathrm{i} & \\
& \mathrm{i} & u+\frac{\mathrm{i}(1-\beta)}{2} & \\
\frac{\mathrm{i}(1-\gamma)}{2} & & & u+\frac{\mathrm{i}(1+\beta)}{2}
\end{matrix}\right).
\end{equation*}
By optimizing both the Bethe roots and the parameters $\beta$ and $\gamma$, we obtained a significantly improved approximation. For the XXZ model, the fidelity can be enhanced to unity, with an error of order of $10^{-13}$. The optimized value of $\gamma$ is very close to 1, consistent with the fact that XXZ spin chains correspond to the 6-vertex model. The improvement can also be seen in models where integrability is broken. In the weak integrability-breaking case, the fidelity is slightly improved towards $1$ at a generic point, and a mysterious improvement for the 1st excited state is observed at the MG point $\lambda = 0.5$ (see Fig.~\ref{fig:4a}). An unexpected phenomenon is observed in the strong integrability-breaking model: the fidelity revives to $1$ when the magnetic field becomes large enough (see Fig. \ref{fig:4b}). We remark that although the 8-vertex ``R‑matrix'' employed here is generally not a solution to the Yang–Baxter equation \footnote{From the perspective of variational method, introducing more parameters generally yields better approximations (with higher computational costs), thus the 8-vertex ansatz adopted here is a naive extension of the original EBA. While the ansatz itself can be chosen arbitrarily, when the R-matrix satisfies the Yang–Baxter equation (e.g. the R-matrix of the XXZ model), it may further offer some physical interpretation to the results.}, an interesting observation is that for the 1st excited state in the strong integrability‑breaking model with $L=10$ and $\lambda\sim 1500$, most components of the Yang–Baxter equation evaluated with the optimized R‑matrix are found to be very close to zero. This suggests that some sector of the free Hamiltonian $H=\sum_{n=1}^L(-1)^n\sigma^z_n$ might also be described by the 8-vertex $R$-matrix. 
\vspace{0.3cm}

\begin{figure}[htbp]
  \centering
  \includegraphics[width=\columnwidth]{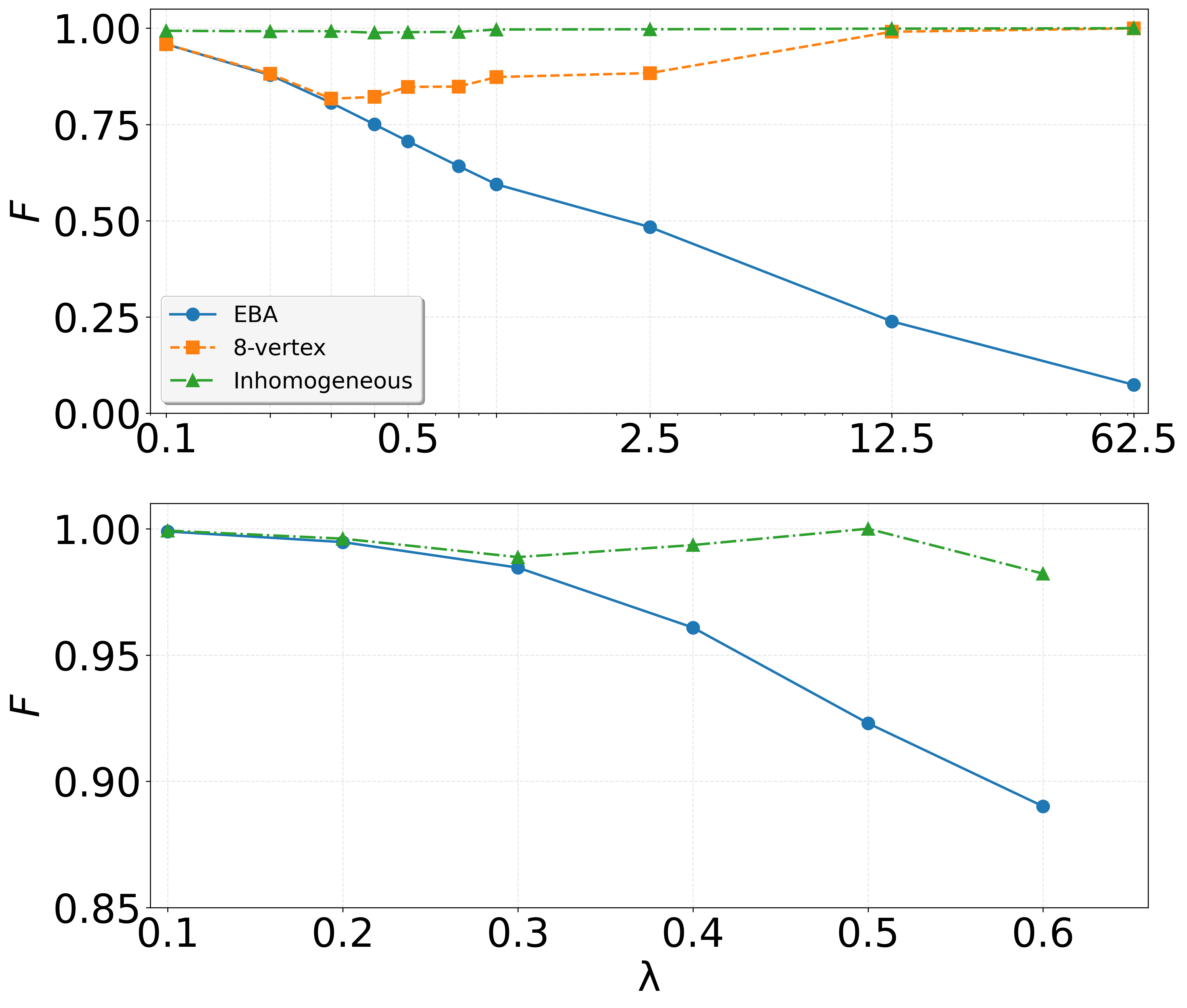}
  \caption{Comparison of ground-state fidelity $F$ for three ansätze in the $L=8$ chain. Upper panel: strong-breaking model; lower panel: weak-breaking model. In the weak-breaking case, the 8-vertex extension offers little improvement over the standard EBA for the ground state (which is difficult to discern in the plot), whereas the inhomogeneous ansatz maintains much better fidelity $F\geq 0.98$ across the full parameter regime in the lower figure.}
  \label{fig:inhom}
\end{figure}

\noindent{\it -Inhomogeneities} The weak integrability-breaking model features long-range interactions beyond the nearest-neighbor level. To approximate long-range interacting spin chains, a natural extension of our ansatz is to further incorporate inhomogeneous parameters $\{\theta_j\}$ into the transfer matrix (see Appendix for the definition) since the inhomogeneous model is intrinsically long-range interacting.
%More precisely, we consider
%\begin{align}
%{\bf T}(u,{\theta_i}):=&{\rm tr}_0\left(R_{0L}(u-\theta_L)R_{0(L-1)}(u-\theta_{L-1})\right.\cr
%&\left.\times \dots R_{01}(u-\theta_1)\right).\label{ihhom-trans}
%\end{align}
Using the effective Bethe ansatz generated from this transfer matrix increases the fidelity $F$ to $F\geq0.99$ for all values of $\lambda$ in the strong integrability-breaking case and $F\geq0.98$ in the region $0\leq \lambda\leq 0.6$ in the weakly breaking model (see Figure \ref{fig:inhom}). This improvement comes at the cost of optimizing both the inhomogeneities and the rapidities, which raises the computational costs. An optimal compromise between computational cost and accuracy is to make a proper choice of the inhomogeneity, which reduces the number of tunable parameters.

\vspace{0.5cm}

\noindent{\bf Discussions and outlook.}
We have introduced an effective Bethe ansatz for quantum systems in the vicinity of an integrable point. The central idea is that sufficiently close to integrability, the functional form of the Bethe wave function remains valid, while the Bethe roots undergo renormalization due to integrability-breaking perturbations. These effective roots are determined by minimizing physically motivated cost functions. The resulting effective Bethe states provide approximations to the exact eigenstates of non-integrable models. With a suitable refinement of the ansatz, the approximation can be remarkably accurate.

The idea of approximating exact eigenstates of non-integrable models by appropriate \emph{off-shell} Bethe states offers numerous avenues for further development. For a given non-integrable model, what is the optimal choice of off-shell Bethe state and cost function? It is plausible that different choices of cost functions may be suited to different purposes. For example, in the XXZ model, one can also start from the XX point to perform the EBA analysis, and for $\Delta\lesssim0.5$, it outperforms the EBA based on the XXX ansatz. Since the XX model maps exactly to free fermions, the EBA of XX version is essentially the same as the Hartree-Fock approximation, and similar ideas have been widely applied in the study of quantum many-body physics by using the Slater-Gaussian state (or its superposition) as a variational ansatz \cite{Bravyi:2016caz,Boutin:2021ngf,Buijsman:2023mbp,Albert:2024pwv}. To gain a deeper understanding, it is necessary to explore more general classes of models and a broader range of physical observables. A natural extension of this work is to apply the method to spin-1 models, such as those arising from deformations of the integrable Takhtajan–Babujian model \cite{Babujian:1982ib,Takhtajan:1982jeo}, which includes the AKLT model \cite{Affleck:1987vf,Affleck:1987cy} as a special case. Progress in this direction has been made and is presented in \cite{zhuohang}.

\begin{comment}
    {\color{blue} Bethe ansatz can be incomplete in certain regimes, e.g., for the XXZ model at roots of unity ($q^N=1$) \cite{Fabricius:2000yx}. Nevertheless, we find empirically that this incompleteness does not obstruct the performance of the EBA method in the low-energy sectors studied in this work. For the antiferromagnetic ground state, EBA accurately describes the Hamiltonian at root of unity $\Delta=-0.5$ perturbed by an XXZ-type next-nearest-neighbor interaction term: e.g. at $L=8$, $\lambda=0.15$, the fidelity yields $0.9862$. While the standard Bethe ansatz fails due to Fabricius–McCoy (FM) strings, descendants in the $\mathfrak{sl}_2$ loop algebra, recent developments such as the constrained rational Q-system \cite{Hou:2023jkr} may offer a promising route to systematically resolve these states and help to study the full Hilbert space under perturbation with EBA.}
\end{comment}

A deeper physical understanding of why the off-shell Bethe ansatz or the factorized scattering picture breaks down for sufficiently strong integrability-breaking interactions would be desirable. In quantum field theory, away from integrability, stable particles may decay, and particle production and annihilation processes become allowed, rendering scattering inelastic. Can such a mechanism also be realized in the spin chain setting? Or are there other underlying mechanisms at play?  Answering these questions would deepen our understanding of integrability breaking and may provide guidance for developing improved ansätze further from integrability.

The algebraic structure of off-shell Bethe states may prove useful for investigating a wider class of physical observables. One important direction concerns dynamical properties, such as quench dynamics and quantum transport. How well does the approximation hold under time evolution? Does the fidelity decay over time or remain robust? Addressing these questions could yield new insights into phenomena such as prethermalization and anomalous transport.

\vspace{1cm}

\noindent{\bf Acknowledgements.} We are grateful for the valuable feedback received during the International Workshop on ``Challenges in Integrability'', where part of this work was presented. The work of Y.J. is supported by National Natural Science Foundation of China through Grant No.12575073. R.Z. is partially supported by National Natural Science Foundation of China through Grant No.12105198. This work is also supported by the NSFC Grant No. 12247103.

\begin{comment}
\begin{figure}[h]
\centering
\includegraphics[width=\columnwidth]{root_shift_figure.pdf}
\caption{Outer-root shift $\delta u_1$ versus $\lambda$: EBA data (circles) versus the first-order integrable long-range prediction $\delta u_1=0.00702\,\lambda$ (dashed, slope from \eqref{eq:LRshift}) and the second-order prediction $\delta u_1=0.00702\,\lambda+0.00639\,\lambda^2$ (solid, \eqref{eq:LRshift2}). The first-order line is the $\lambda\to0$ slope $\delta u_1/\lambda$; the second-order parabola captures most of the bending.}
\label{fig:LRshift}
\end{figure}
\end{comment}

\appendix

\section{Algebraic Bethe ansatz}
This appendix provides a concise review of the algebraic Bethe ansatz (ABA) for the Heisenberg spin chain. For a more detailed introduction, we refer to \cite{Faddeev:1994nk,Slavnov:2018kfx}.
\vspace{0.3cm} 

\noindent{\it XXX spin chain.} We begin with the 6-vertex $R$-matrix, which acts on the tensor product space $\mathbb{C}_a^2\otimes\mathbb{C}_b^2$ and is given by
\begin{align}
R_{ab}(u)=\begin{pmatrix} u+\mathrm{i} & 0 & 0 & 0\\ 0& u & \mathrm{i}& 0\\ 0 &\mathrm{i} & u & 0\\ 0&0&0&u+\mathrm{i}\end{pmatrix}
\end{align}
The central object in the ABA framework is the monodromy matrix, defined as
\begin{align}
\mathbf{M}_a(u)\equiv R_{aL}(u-\theta_L)\ldots R_{a1}(u-\theta_1)
\end{align}
where $L$ is the length of the spin chain and  `$a$' denotes a two-dimensional auxiliary space, the parameters $\{\theta_i\}$ are called inhomogeneities. For the special case $\theta_1=\theta_2=\ldots=\theta_L$, we recover the homogeneous model. In this auxiliary space, the monodromy matrix can be expressed as a $2\times 2$ matrix
\begin{align}
\mathbf{M}_a(u)=\begin{pmatrix} \mathbf{A}(u) & \mathbf{B}(u)\\ \mathbf{C}(u) &\mathbf{D}(u) \end{pmatrix}_a\,.
\end{align}
The transfer matrix, which serves as the generating function for the conserved charges, is obtained by taking the trace over the auxiliary space:
\begin{align}
\mathbf{T}(u)=\text{Tr}_a\mathbf{M}_a(u)=\mathbf{A}(u)+\mathbf{D}(u)\,,\label{def-trans}
\end{align}
and has the fundamental property that $[\mathbf{T}(u),\mathbf{T}(v)]=0$.
The Hamiltonian of the XXX chain is then derived from the transfer matrix of the homogeneous model (with $\theta_j=\tfrac{\mathrm{i}}{2}$) as
\begin{align}
H_{\text{XXX}}\propto\left. \frac{\mathrm{d}}{\mathrm{d}u}\log\mathbf{T}(u)\right|_{u=\mathrm{i}/2}
\end{align}
Consequently, eigenvectors of the transfer matrix are also eigenvectors of the Hamiltonian. Such eigenstates can be constructed by repeatedly acting with the $B$-operator on the pseudovacuum state $|\Omega\rangle=|\uparrow^L\rangle$
\begin{align}
|\{u_j\}\rangle=\mathbf{B}(u_1)\ldots\mathbf{B}(u_N)|\Omega\rangle
\end{align}
For this state to be an eigenstate of the transfer matrix, the complex parameters $\{u_j\}$ must satisfy the Bethe ansatz equations (BAE)
\begin{align}
\left(\frac{u_j+\tfrac{\mathrm{i}}{2}}{u_j-\tfrac{\mathrm{i}}{2}} \right)^L=\prod_{k=1\atop k\ne j}^N\frac{u_j-u_k+\mathrm{i}}{u_j-u_k-\mathrm{i}},\quad j=1,\ldots,N.
\end{align}
A Bethe state whose rapidities satisfy the BAE is referred to as on-shell; otherwise, it is called off-shell.

\vspace{0.3cm} 

\noindent{\it XXZ spin chain.} The XXZ spin chain is defined by the Hamiltonian
\begin{align}
H_{\text{XXZ}}=\sum_{n=1}^L\left(\sigma_n^x\sigma_{n+1}^x+ \sigma_n^y\sigma_{n+1}^y+\Delta \sigma_n^z\sigma_{n+1}^z\right)
\end{align}
where $\Delta$ is the anisotropic parameter and $\Delta=1$ corresponds to the XXX spin chain. The XXZ spin chain can also be solved by ABA, with the following $R$-matrix
\begin{align}
R_{ab}(u)=\begin{pmatrix} \sinh(u+\eta) & 0 & 0 & 0\\ 0& \sinh u & \sinh\eta & 0\\ 0 &\sinh\eta & \sinh u & 0\\ 0&0&0&\sinh(u+\eta)\end{pmatrix}
\end{align}
with an additional parameter $\eta$. This parameter is related to the $q$-deformation parameter and anisotropy by
\begin{align}
q=e^{\eta},\qquad \Delta=\frac{1}{2}(q+q^{-1})=\cosh\eta\,.
\end{align}
The rest of the construction follows the same way as the XXX spin chain and will not be repeated here.

\section{Numerical Optimization of Effective Bethe Roots}

This appendix briefly outlines the numerical algorithm employed to optimize the parameters of the effective Bethe ansatz. The goal is to find, through variational optimization of ${\cal H}_n$, a set of complex parameters, including the effective Bethe roots $\{u_j\}$ and other shared algebraic parameters such as $\beta,\gamma$, and inhomogeneous parameters $\{\theta_j\}$ potentially existing when the ansatz is further improved. 

The algorithm we use is implemented in Python. The core computations rely on the following libraries:
\begin{itemize}
    \item {\bf SciPy}: Used for sparse matrix representation of the Hamiltonian and for exact diagonalization to obtain benchmark energies and states to be compared with. Its module, scipy.optimize.minimize, provides an efficient optimizer, {\bf L-BFGS-B} optimizer, a quasi-Newton method suitable for bound-constrained, medium-dimensional problems.
    \item {\bf PyTorch}: Used to implement the construction of the effective Bethe state and the computation of the energy expectation value. Its automatic differentiation capability is leveraged to compute the exact gradient of the loss function with respect to all $2M\ (+\alpha)$ real parameters, which is crucial for efficient optimization convergence. For relatively large systems (e.g., $L\sim 20$), where the Hilbert space dimension grows exponentially, the vectorized operations and reverse-mode automatic differentiation in PyTorch provide a significant speed advantage over manual gradient implementation or finite-difference methods.
\end{itemize}

To ensure robust convergence and avoid suboptimal local minima, the procedure employs the following initialization and search strategies: 
\begin{itemize}
    \item Random Initialization: The real and imaginary parts of effective Bethe roots $\{u_j\}$ are independently sampled from a normal distribution with mean $0$ and standard deviation $10$ as the initial guess.
    \item Parallel Attempts: The program performs multiple independent optimization runs (e.g., 30), each with a different random initial point. Each optimization is subject to a maximum number of iterations (e.g., 6,000) and convergence tolerances based on function value and gradient (ftol=1e-12, gtol=1e-10).
    \item Selection of Best Result: After each run, the final energy expectation and the fidelity (overlap) with the exact ground and first excited states (obtained via exact diagonalization) are recorded. The solution with the lowest energy among all attempts is selected as the final effective Bethe state.
    \item {\bf High-Quality Initialization via BAE}: For systems near the integrable point ($\lambda\ll 1$), a highly effective initialization is obtained by first numerically solving the BAE for the underlying integrable XXX model ($\lambda=0$) for the target sector. The resulting set of Bethe roots $\{u^{(0)}_j\}$ provides an excellent initial guess for the variational parameters $\{u_j\}$, and this strategy places the optimizer in the vicinity of the global minimum, dramatically accelerating convergence.
\end{itemize}

\begin{figure}[htbp]
  \centering
  \includegraphics[width=\columnwidth]{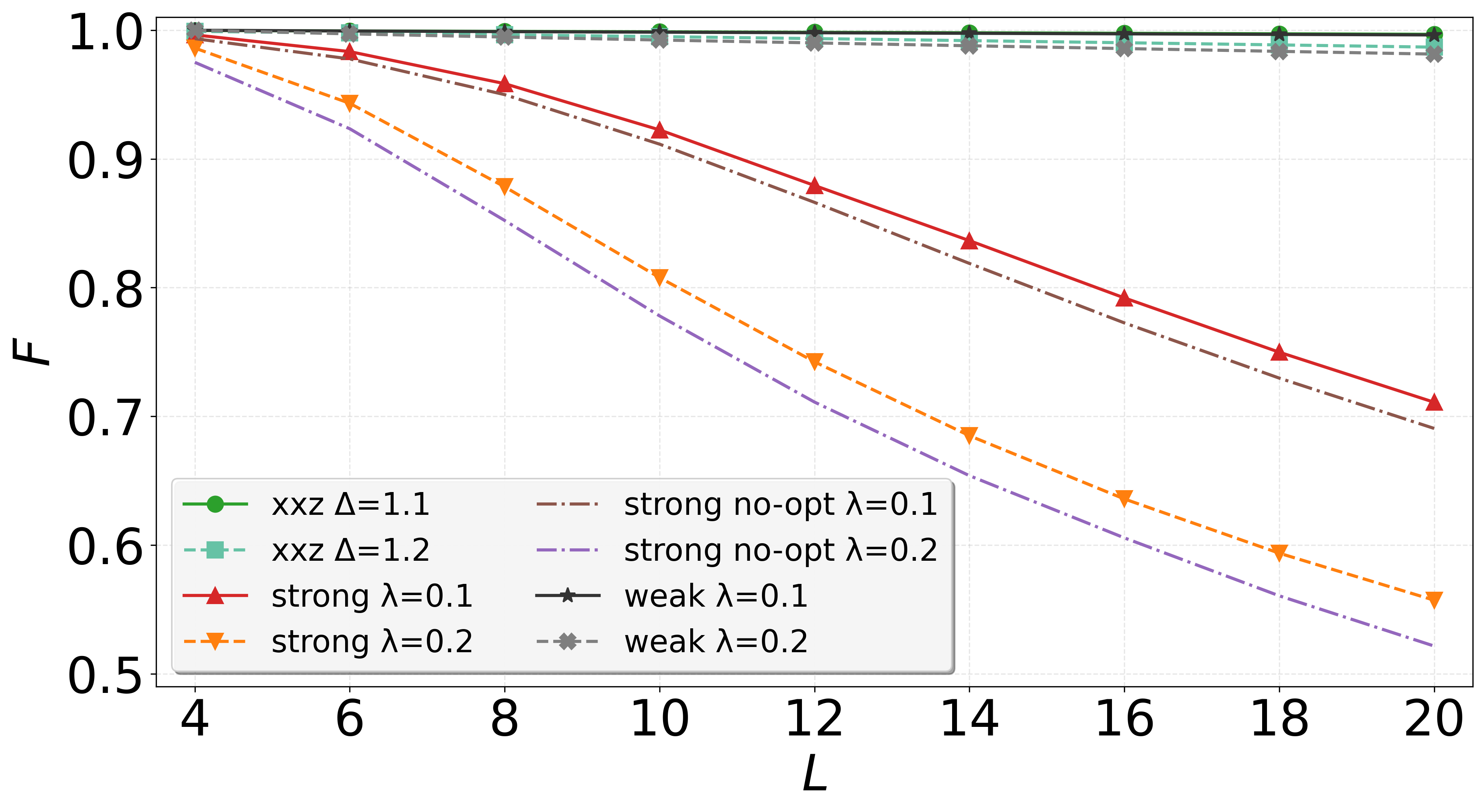}
  \caption{Ground-state fidelity $F$ versus system size $L$ for the XXZ chain ($\Delta=1.1,1.2$) and the integrability-breaking models (strong and weak, $\lambda=0.1,0.2$). The curves are fitted with $F=a+bL^1+cL^2$. The coefficients $(a,b,c)$ are found to be: 
    XXZ$_{\Delta=1.1}$: (1.000389, -1.47e-4, -2e-6), \hskip 2cm
    XXZ$_{\Delta=1.2}$: (1.001526, -5.66e-4, -8e-6), \hskip 2cm
    strong$_{\lambda=0.1}$: (1.041342, -7.968e-3, -4.47e-4), \hskip 2cm
    strong$_{\lambda=0.2}$: (1.136773, -3.7269e-2, 3.97e-4), \hskip 2cm
    weak$_{\lambda=0.1}$: (1.001017, -2.76e-4, 2e-6), \hskip 2cm
    weak$_{\lambda=0.2}$: (1.005148, -1.364e-3, 9e-6). \hskip 3cm
  All fittings achieve $R^2>0.996$. The coefficients of $L^2$-term are almost negligible except for the strong breaking model. The no-opt curves give the fidelity from the XXX Bethe states without optimization of Bethe roots.
  }
  \label{fig:eba_vs_L}
\end{figure}

This pipeline successfully produced high-fidelity approximations for systems near the integrable point as discussed in the main text, and in practice, we have checked its effectiveness up to $L=24$ (we managed to run our program for $L=26$, but ED was not available on the same computer for comparison). Figure~\ref{fig:eba_vs_L} shows the fidelity as a function of system size $L$. 

\end{CJK*}

\bibliography{yunfeng} 
\bibliographystyle{utphys}
\end{document}